\documentclass[runningheads]{llncs}
\usepackage{lmodern} 
\usepackage{diagbox}
\usepackage[breaklinks=true]{hyperref}
\usepackage{textcomp}
\usepackage{makecell}

\usepackage{subcaption}
\usepackage{mathtools}
\usepackage{amsmath}
\usepackage{amssymb}
\usepackage{multirow}
\usepackage{rotating}
\usepackage{multicol}
\usepackage{paralist}
\usepackage{amsfonts}
\usepackage{bbm}
\usepackage{graphics}      
\usepackage{txfonts}
\usepackage{color}
\usepackage{colortbl}

\usepackage{booktabs}
\usepackage{textcomp}
\usepackage{graphicx}
\graphicspath{{figures/}}

\begin{document}
\title{Inside the x-rated world of ``premium'' social media accounts}

\author{
Nikolaos Lykousas\inst{1}
\and
Fran Casino\inst{1}
\and
Constantinos Patsakis\inst{1,2}
}

\institute{
University of Piraeus, Greece. \email{\{nlykousas,fran.casino,kpatsak\}@unipi.gr}\\
\and
Athena Research Center, Greece
}

\maketitle 
\begin{abstract}
During the last few years, there has been an upsurge of social media influencers who are part of the adult entertainment industry, referred to as \textit{Performers}.
To monetize their online presence, Performers often engage in practices which violate community guidelines of social media, such as
selling subscriptions for accessing their private ``premium'' social media accounts, where they distribute adult content.
In this paper, we collect and analyze data from FanCentro, an online marketplace where Performers can sell adult content and subscriptions to private accounts in platforms like Snapchat and Instagram.
Our work aims to shed light on the semi-illicit adult content market layered on the top of popular social media platforms and its offerings, as well as to profile the demographics, activity and content produced by Performers. 

\end{abstract}
\section{Introduction}
In the world of social media, content creators play a central role in shaping a global online culture. The content creators who raise in popularity can attain the status of online micro-celebrities, and they are commonly characterized as \textit{influencers} \cite{khamis2017self}.
The main objective of influencers is to produce digital content which attracts users' attention and rapidly gains popularity, often becoming `viral', in platforms such as Instagram and YouTube \cite{gomez2019digital,nandagiri2018impact}. In this regard, influencers leverage focused visual content and targeted communication techniques to capture and sustain the attention of social media users, thus building large follower bases and attaining organic social reach.
Social media content creators can thus monetize their reach in various ways, such as using word-of-mouth marketing techniques and promoting brands and campaigns \cite{terranova2012attention,lou2019influencer}.

One of the most prevalent strategies employed by influencers %as a means
to entice followers towards heightened forms of emotional engagement is sexualized labour \cite{drenten2020sexualized}. Posting sexualized images in social media is a popular form of self-presentation for young adults \cite{daniels2016sexiness,baumgartner2015sexual,van2015exploring,van2017sexy}, and it is outlined as the core tactic to attract followers for a particular type of influencers, which are categorized as \textit{``Performers''} in \cite{drenten2020sexualized}. 

This category of influencers includes adult performers/entertainers, sex workers and models. In all cases, after building an audience in mainstream social media, Performers redirect their followers to external outlets for purchasing exclusive content, often pornographic in nature. %more overtly sexual
Notable examples of such outlets are platforms like OnlyFans\footnote{\url{https://onlyfans.com}} (effectively an `adult' version of Instagram), and ``premium'' Snapchat accounts, offering a lucrative income stream for Performers looking to monetize their online presence \cite{wired2019xrated}. 
%The most common forms of payment models for such services are subscriptions or so-called `lifetime access'. Referring to recurring and one-off payments to access new content, respectively.
For social media platforms like Snapchat, the community guidelines\footnote{\url{https://www.snap.com/en-US/community-guidelines}} explicitly \textit{prohibit accounts that promote or distribute pornographic content}. Nonetheless, it has been shown that community guidelines cannot be effectively enforced to ban adult content in social media \cite{lykousas2018adult}. %move this to discussion  ?
As such, Performers who systematically violate community guidelines by posting overtly sexual content, have to use external means for managing transactions with their client base, as well as maintaining their digital presence in multiple social outlets, in case their accounts get suspended. % in mainstream social platforms

In this paper, we analyze data collected from \textit{FanCentro}\footnote{\url{https://fancentro.com}}, a platform where Performers can monetize their fan base via selling subscriptions to their private social media accounts. Additionally, FanCentro enables Performers to directly sell private content through a media feed, as well as chatting functionality between Performers and their subscribers. As a requirement for opening an account in FanCentro, Performers have to provide a digital copy of government-issued ID for age verification purposes. After this verification step, FanCentro, for a fraction of the paid subscriptions, handles all of the necessary transactions and administrative activities. 

There are two main reasons we chose FanCentro over other similar platforms such as OnlyFans, which have gained wide mainstream media attention \cite{elle2020onlyfans}. First, its primary focus is selling access to ``premium'' accounts in social platforms which, strictly, are not content marketplaces (i.e. Snapchat and Instagram). Second, FanCentro website provides a complete listing\footnote{In contrast, OnlyFans platform does not have such functionality.} of Performer profiles, enabling us to collect data without having to employ sampling techniques which could potentially bias our findings. % allowing us to obtain an unbiased dataset without having
Our work aims to shed light on the mechanics of the semi-illicit industry of premium social media subscriptions and services offered by Performers, in the context of adult content marketplaces such as FanCentro.
% \todo[inline]{FC minimise description of fancentro in the intro and leave it for later. Instead, focus on more general numbers if you have them and global statistics of this kind of "illegal practice" }

% \section{Related Work} 
% % \todo[inline]{FC I understand that you want to go to the point and that there is maybe no need for this, but any reviewer can state the novelty or can properly evaluate the work without "ground truth". Therefore, a related work section is needed. It can be brief if there is not too much on this ofc. Just state what are other tools or apps maybe and what research or similar stuff has been done. IF not, state it.}
% \section{Method}

% Step by step of the procedures that you will do. A pic would help.  You are going to collect data, analyse it and see the outcomes... Try to sell that your working procedure is appropriate. You can also do a rationale section, analysing who are the actors and which is the data of the system and what you want to analyse.  Describe what is fancentro here in detail also .

% In general, the procedure or methods that you will follow.

\begin{figure}[!ht]
      \centering
   \includegraphics[width=0.769\columnwidth]{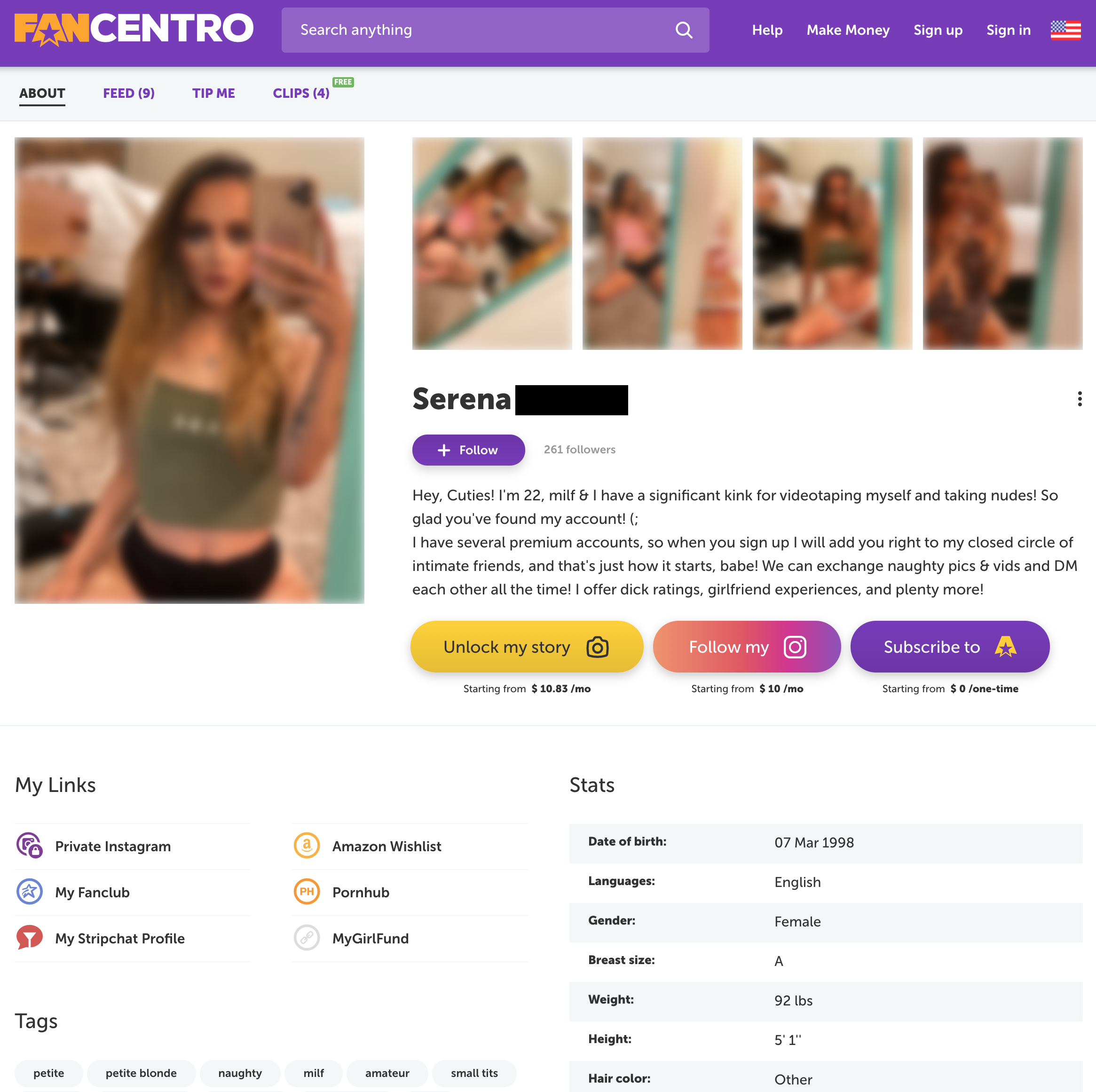}
    \caption{An example Performer's profile page in FanCentro.}
    \label{fig:fc_screen}
\end{figure}
\section{Data Collection}
We constructed a \emph{complete} dataset with the profiles of Performers registered in FanCentro as of April 5th, 2020. In Figure~\ref{fig:fc_screen}, we provide an illustrative example of a Performer's profile page. We note that only Performers have public profiles and can post content, while regular users/subscribers can only interact with Performers (i.e. follow, message, like/comment to their posts) and not other users. 
In total, we collected the profile attributes, published content metadata, and offered products for $16,488$ users.
For this, we created a crawler which consumes the API used by FanCentro's website, enabling us to collect the relevant data. %public
%attributes of Performers' profiles and the services they sell.
Despite the ``public'' nature of collected information, we follow Zimmer's approach \cite{zimmer2010but}. In this regard, the data remains anonymized during all the steps of our analysis, and we report only aggregate findings.

\begin{figure}[!hb]
      \centering
   \includegraphics[width=1.02\columnwidth]{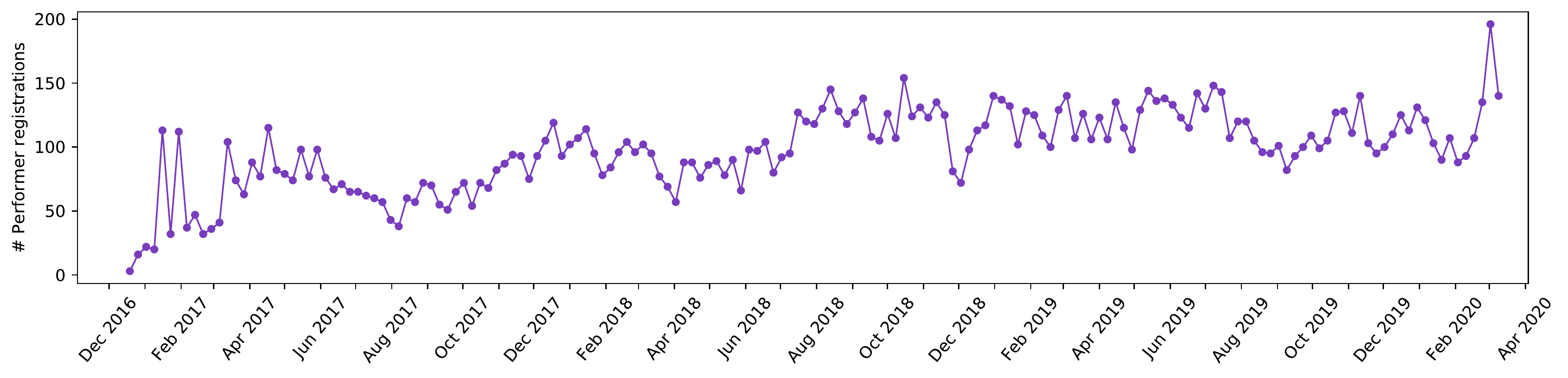}
    \caption{Weekly registrations}
    \label{fig:regs}
\end{figure}

In order to measure the activity in FanCentro in terms of new registrations of Performers, in Figure~\ref{fig:regs}, we plot the number of accounts created each week since the launch of the platform. From January of 2017 (FanCentro launch), the weekly registrations show an increasing trend until a peak was reached in November of 2018. Since then the registration rate has been generally sustained, until we observe a spike in registrations the last week of March 2020, followed by the first week of April 2020, with 196 and 161 new users, respectively.
This sharp increase in new users towards the end of March 2020 is also reflected in other similar sites, and it can be linked to the coronavirus pandemic, the consequent lockdowns, and its implications for sex work \cite{nytimes2020sexwork,wired2020corona}.  % the transformation of sex work to a ``gig economy''

\section{Results and Discussion}
\subsection{Characterizing Performers}

% \begin{table}[!ht]
% \centering
% \begin{tabular}{c c c c | r }
% \toprule
% \midrule
% \multirow{2}[3]{*}{\thead{Sexual \\ Orientation}} & \multicolumn{4}{c}{\thead{Sexual Identity}}\\ 
% \cmidrule(rl){2-5}
% & Female  & Male & Trans & Total  \\ 
% \cmidrule(r){1-1}\cmidrule(l){2-5}
% \multicolumn{1}{l}{Bisexual}& 2486 & 52 & 41 & 2579   \\
% \multicolumn{1}{l}{Gay}& 202 & 34 & 5 & 241  \\
% \multicolumn{1}{l}{Straight} & 3544 & 141 & 27 & 3712  \\
% \multicolumn{1}{l}{Trans} & 18 & 4 & 44 & 66  \\
% \midrule
% \multicolumn{1}{l}{Total} & 6250 & 231 & 117 & 6598  \\
% \midrule
% \bottomrule
% \end{tabular}
% \caption{Sexual identity and orientation }
% \label{tab:demographics}
% \end{table}

% \begin{figure}[!ht]
% \centering
% \includegraphics[width=0.5\columnwidth]{ages.pdf}
% % \captionsetup{type=figure}
% \caption{Age Distribution}
% \label{fig:ages}
% \end{figure}

\begin{minipage}{\textwidth}
\begin{minipage}[b]{0.49\textwidth}
\centering
\begin{tabular}{c c c c | r }
\toprule
\midrule
\multirow{2}[3]{*}{\thead{Sexual \\ Orientation}} & \multicolumn{4}{c}{\thead{Sexual Identity}}\\ 
\cmidrule(rl){2-5}
& Female  & Male & Trans & Total  \\ 
\cmidrule(r){1-1}\cmidrule(l){2-5}
\multicolumn{1}{l}{Bisexual}& 2486 & 52 & 41 & 2579   \\
\multicolumn{1}{l}{Gay}& 202 & 34 & 5 & 241  \\
\multicolumn{1}{l}{Straight} & 3544 & 141 & 27 & 3712  \\
\multicolumn{1}{l}{Trans} & 18 & 4 & 44 & 66  \\
\midrule
\multicolumn{1}{l}{Total} & 6250 & 231 & 117 & 6598  \\
\midrule
\bottomrule
\end{tabular}
% \captionsetup{type=table}
\captionof{table}{Sexual identity and orientation }
\label{tab:demographics}
\end{minipage}
\hfill
\begin{minipage}[b]{0.49\textwidth}
\centering
\includegraphics[width=1.02\columnwidth]{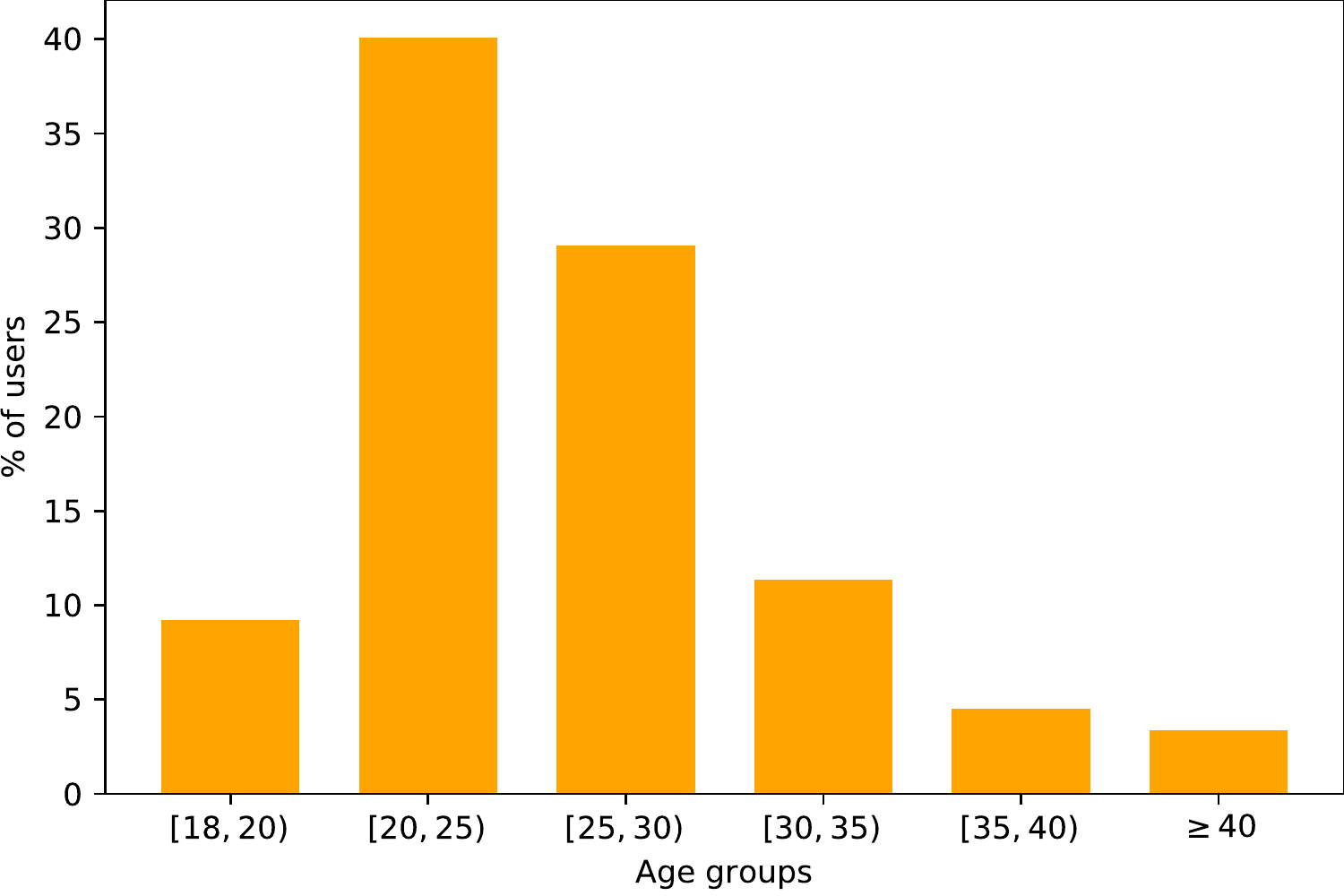}
% \captionsetup{type=figure}
\captionof{figure}{Age Distribution}
\label{fig:ages}
\end{minipage}
\end{minipage}
\\
\\

\begin{figure*}[!ht]
  \centering
  \begin{subfigure}[t]{0.49\textwidth}
      \centering
    \includegraphics[width=\textwidth]{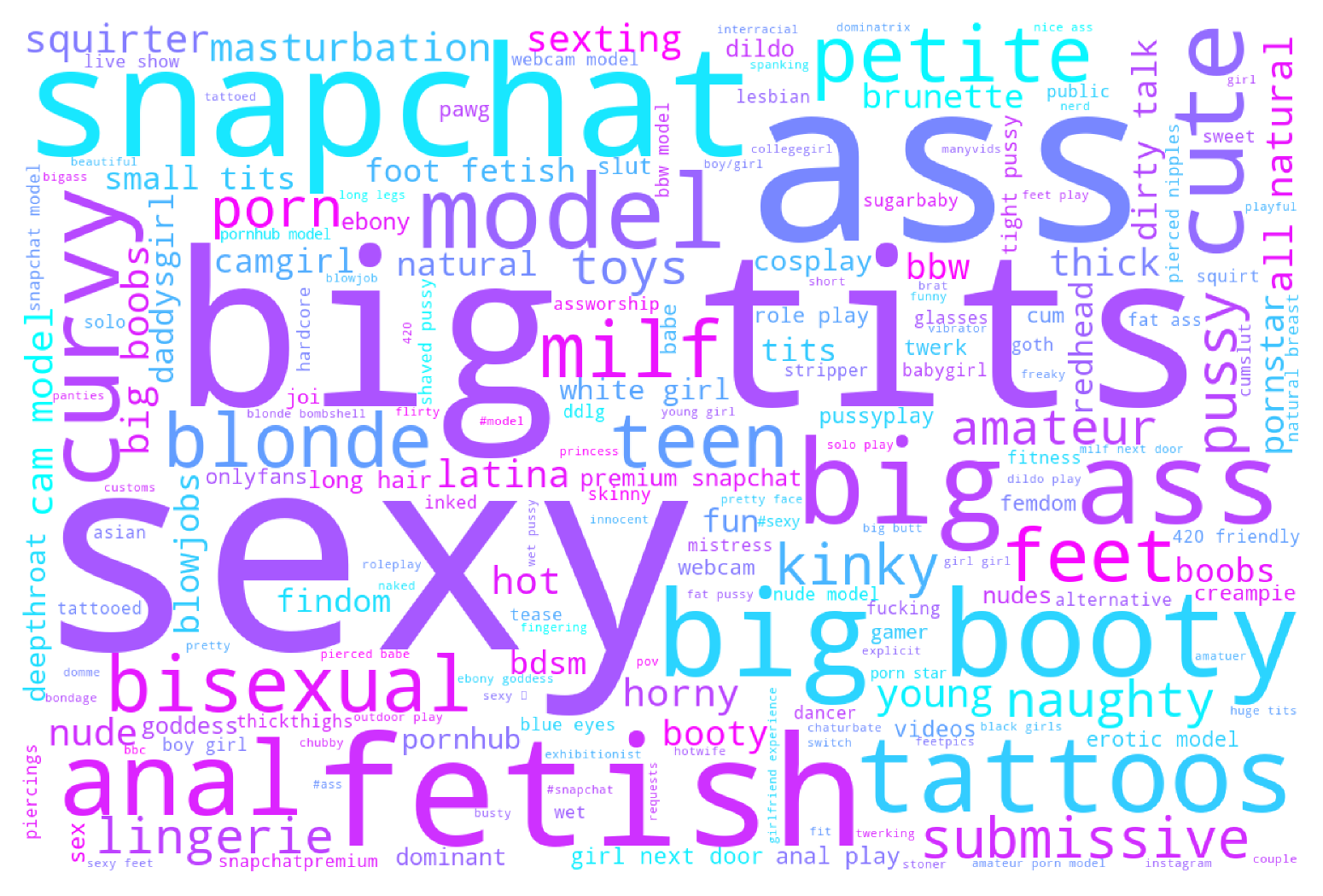}
    \caption{Profile Tags}
    \label{fig:tags}
  \end{subfigure}
  \begin{subfigure}[t]{0.49\textwidth}
      \centering
    \includegraphics[width=\textwidth]{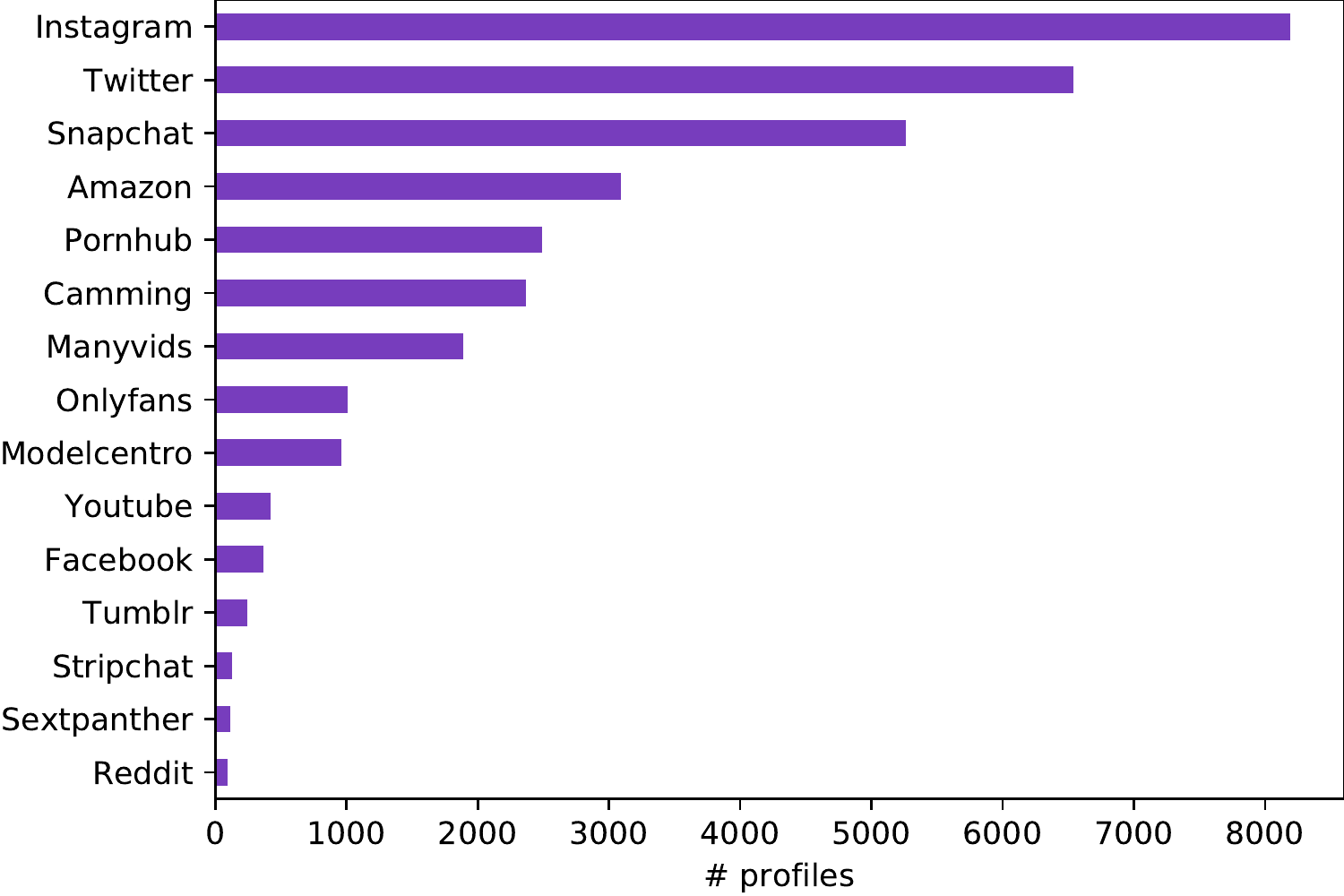}
    \caption{Links}
    \label{fig:links}
  \end{subfigure}  
  \caption{Descriptive characteristics of Performers profiles: tags and links to external sites.}
  \label{fig:metadata}
\end{figure*}

In this section, we study the collected profiles in terms of characterizing attributes. This includes self-reported demographic information (i.e. sexual identity and orientation, age), descriptive tags, and external links to other sites, as provided by Performers. In Table \ref{tab:demographics}, we report the number of profiles per sexual identity and orientation. Notably, $9,879$ profiles did not include this information. Nevertheless, after analyzing the rest of the profiles, we can conclude that the majority of Performers identify as straight females. In Figure \ref{fig:ages}, we depicted the age distribution for the profiles containing the birthdate attribute ($4,526$ profiles). We observe that the most common age group is 20-25 years ($1,857$ profiles), followed by 25-30 ($1,347$ profiles). The latter means that the 70\% of Performers who reported their birthday are within the age bracket of 20 to 30 years.
The next step of our analysis focused on the tags used by the Performers. In this regard, Figure \ref{fig:tags} shows a WordCloud representation of the most frequent tags used by Performers (found in $4,558$ profiles). We observe that they mostly include pornographic terms, with ``sexy'' and ``ass'' being the most popular ($1,472$ and $928$ occurrences, respectively). The outcomes of the analysis of the external links are depicted in In Figure~\ref{fig:links}. We can observe that the most common external links from the profiles collected in our dataset are Instagram and Twitter, closely followed by public Snapchat accounts. This indicates that Performers orchestrate their online presence across multiple social outlets, enabling them to reach and engage a diverse audience.  Moreover, Amazon wish lists, webcam modelling (\textit{``camming''}) platforms \cite{henry2017always} and porn sites have a relevant representation.

% \begin{figure*}[ht]
%   \centering
%   \begin{subfigure}[t]{0.49\textwidth}
%       \centering
%     \includegraphics[width=\textwidth]{tag_cloud.pdf}
%     \caption{Revenue}
%     \label{fig:tags}
%   \end{subfigure}
%   \begin{subfigure}[t]{0.49\textwidth}
%       \centering
%     \includegraphics[width=\textwidth]{links.pdf}
%     \caption{Followers}
%     \label{fig:links}
%   \end{subfigure}  
%   \caption{Histograms of vents per emotion distributed per emotion categories. Disabled emotions are in gray.}
% \end{figure*}

% \begin{minipage}[!t]{\textwidth}
% \begin{minipage}[b]{0.49\textwidth}
% \centering
% \centering
% \includegraphics[width=\columnwidth]{tag_cloud2.pdf}
% % \captionsetup{type=figure}
% \captionof{figure}{Tags}
% \label{fig:tags}
% \end{minipage}
% \hfill
% \begin{minipage}[b]{0.49\textwidth}
% \centering
% \includegraphics[width=\columnwidth]{links.pdf}
% % \captionsetup{type=figure}
% \captionof{figure}{Links}
% \label{fig:links}
% \end{minipage}
% \end{minipage}

\subsection{Exploring the supply and demand}
In order to get an insight into the activities performed in FanCentro, we analyze the metadata information related to the collected profiles, including the amount of funds payable to the Performer in the next payout (revenue), the followers and the content posted by Performers.\footnote{
Revenue is personal in nature and is normally visible only via the dashboard of each Performer. We have contacted FanCentro regarding this matter, and it has been removed from the data delivered via the public API.}

\begin{figure*}[!ht]
  \centering
  \begin{subfigure}[t]{0.325\textwidth}
      \centering
    \includegraphics[width=\textwidth]{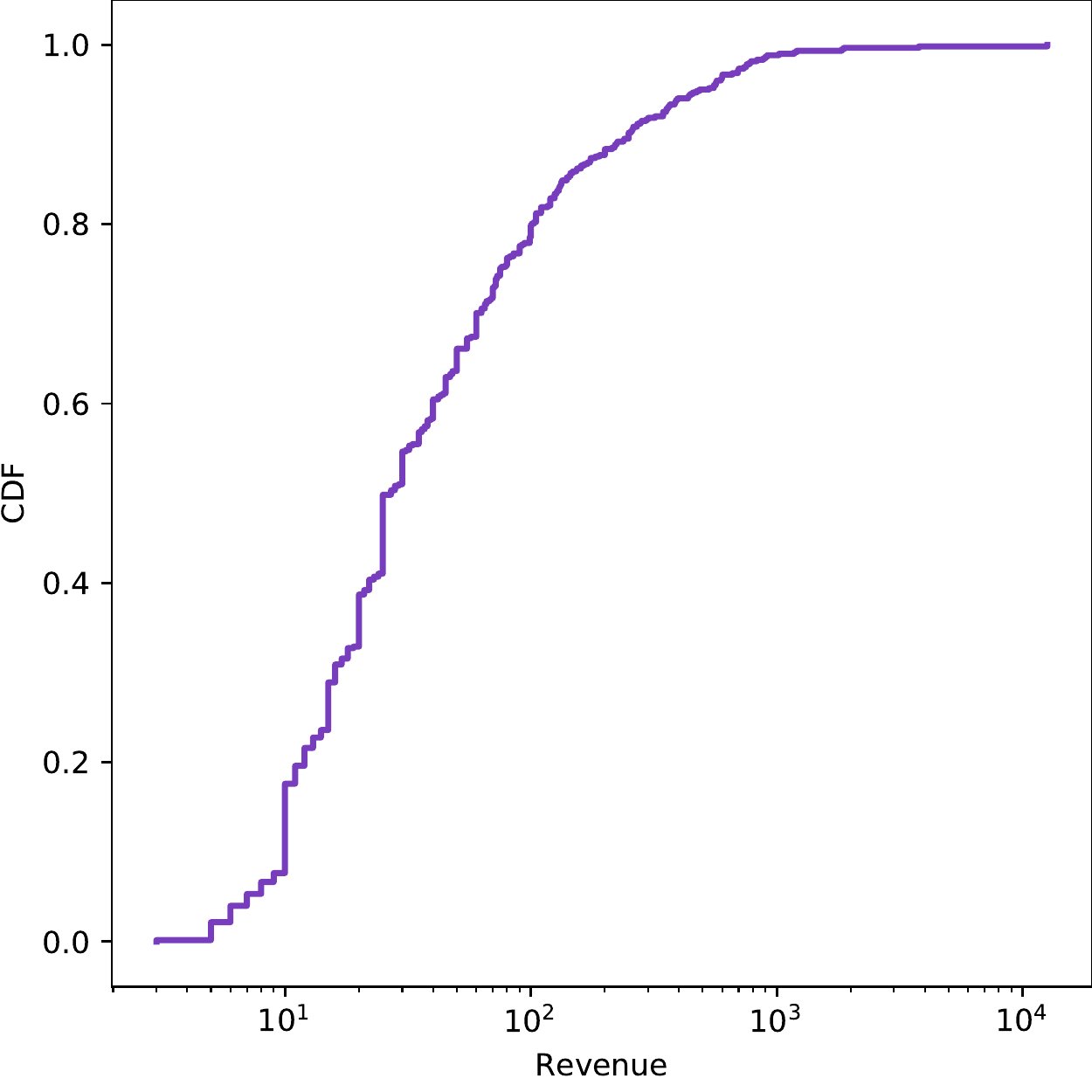}
    \caption{Revenue}
    \label{fig:revenue}
  \end{subfigure}
  \begin{subfigure}[t]{0.325\textwidth}
      \centering
    \includegraphics[width=\textwidth]{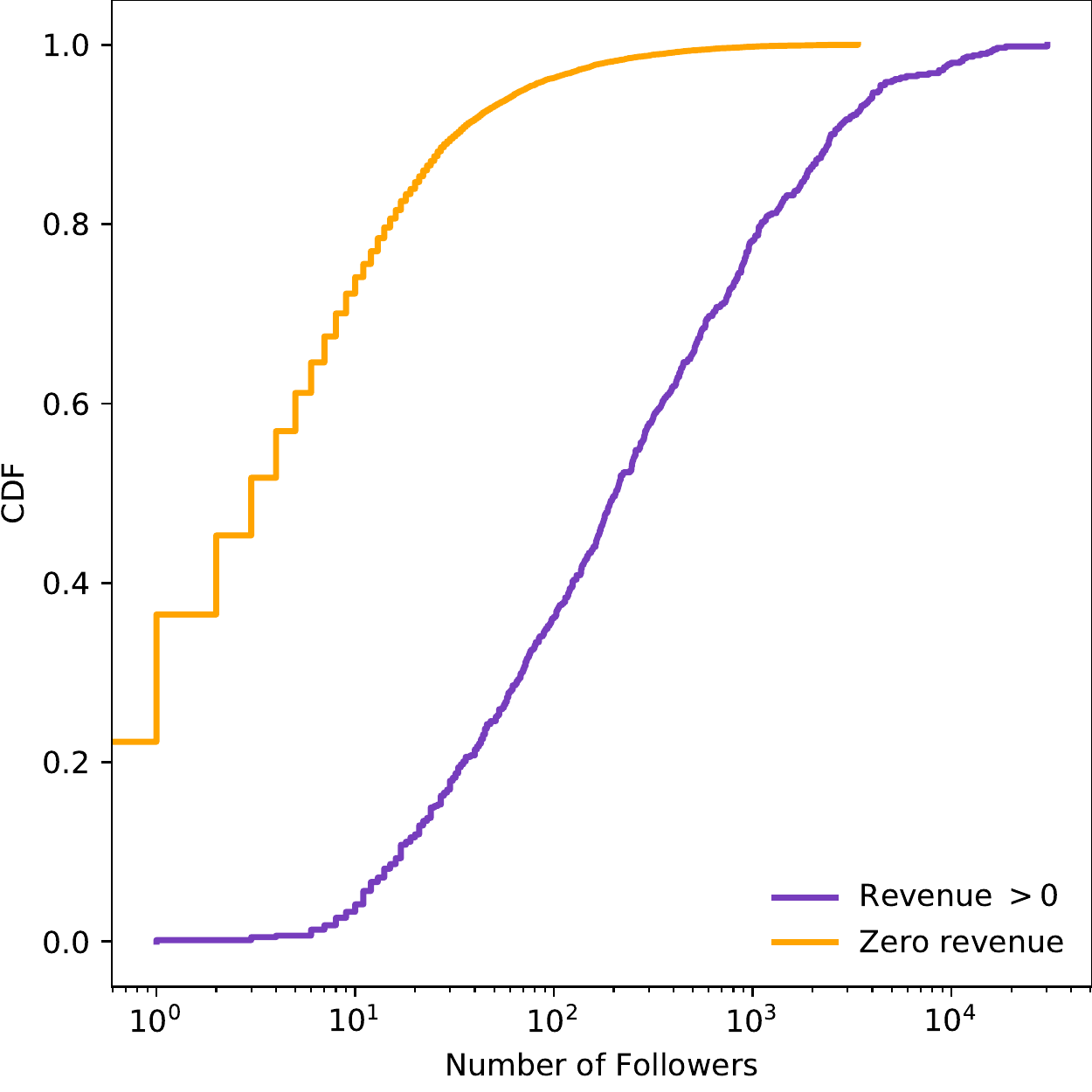}
    \caption{Followers}
    \label{fig:followers}
  \end{subfigure}  
  \begin{subfigure}[t]{0.325\textwidth}
      \centering
    \includegraphics[width=\textwidth]{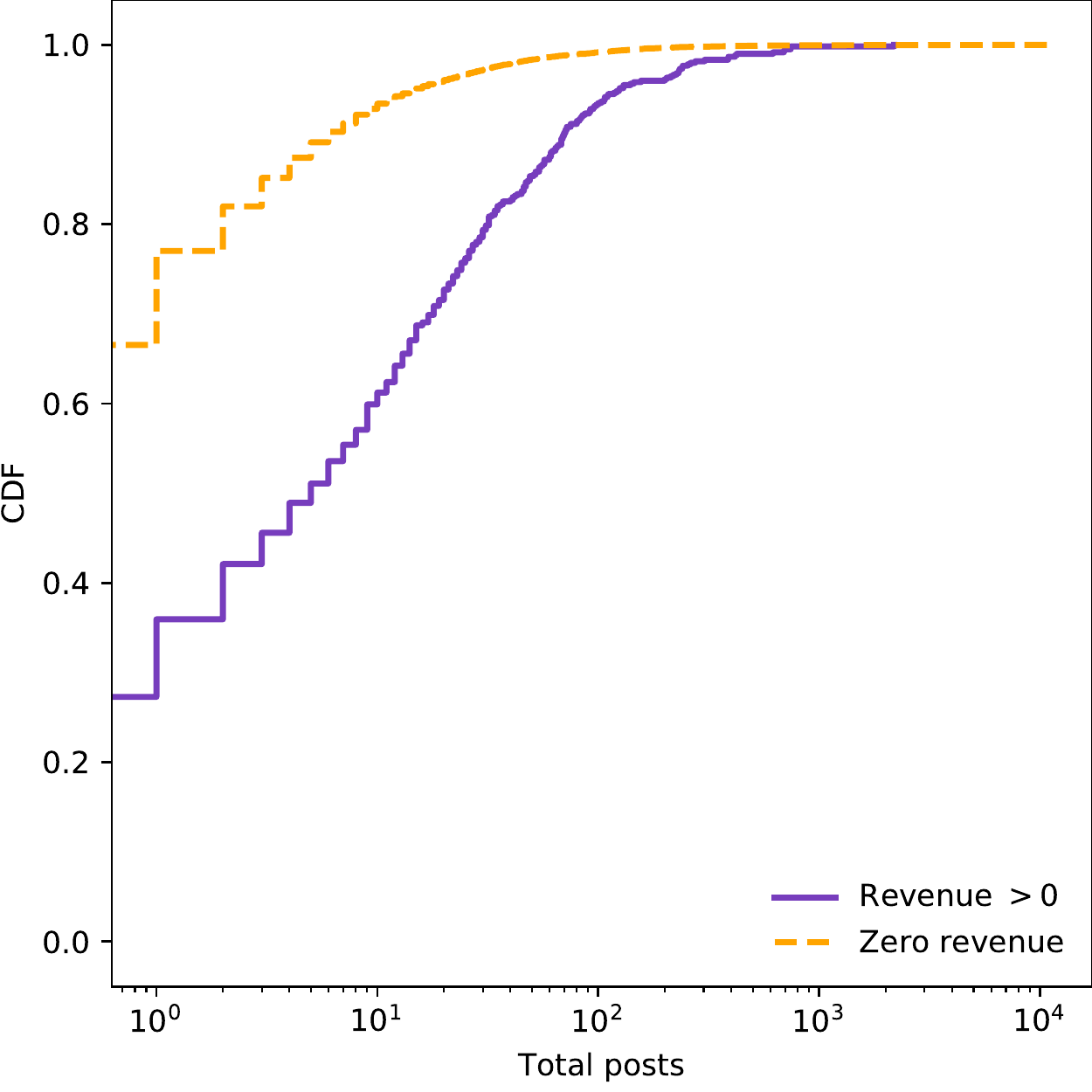}
    \caption{Posts}
    \label{fig:posts}
  \end{subfigure}  
  \caption{Cumulative distribution functions (CDFs) of (non-zero) revenue, number of followers and posts.}
  \label{fig:cdfs}
\end{figure*}

%\subsection{Revenue}
The \textit{revenue} reflects the monetary sum of recurring sales (i.e. subscriptions) at crawl time, plus any income from one-off payments (including gratuity/tips, video clip sales and `lifetime access' services) that are on hold by FanCetro until the next payout to the content creator.\footnote{FanCentro pays Influencers once a week after two weeks of the revenue generation date according to the license agreement.} %as per \url{https://fancentro.com/agreement}}. 
Provided the dynamic nature of subscriptions and content produced by Performers, revenue is a quantity that fluctuates due to a variety of reasons, including cancellation of subscriptions, chargebacks, external factors governing Performers' popularity, etc.
To assess the extent to which the revenue fluctuates over time, we use a snapshot of FanCentro profiles that we collected on March 2nd, 2020. To this end, a two-tailed Kolmogorov–Smirnov test was used, revealing no significant differences in Performers' revenues between two consecutive months ($p=0.44$). We found that the revenue distribution is extremely skewed, with the overwhelming majority of the Performers (96.4\%) generating zero revenue within the aforementioned period. %the two weeks before our data collection.
In Figure~\ref{fig:revenue}, we plot the revenue cumulative distribution function (CDF) for the 602 revenue-earning Performers (3.6\% of profiles). We observe that 80\% are below the minimum payout threshold of 100 USD\footnote{\url{https://centroprofits.com/faq}}, meaning that only a negligible fraction of the performers in our dataset (0.8\% approx.) would be certain to receive income by FanCentro during the next payout. Nonetheless, the revenues for the period between 23 March - 5 April 2020 period reach up to $12,615$ USD. In total, the gross earnings of Performers amount to $73,607$ USD for the payout period captured in our dataset.

Next, in Figure~\ref{fig:followers}, we show the CDF of the number of followers. Contrary to the revenue, 78.5\% of the profiles have followers. However, the revenue-generating Performers have up to two orders of magnitude more followers than the rest. The statistical significance of this difference was also confirmed by a two-tailed Kolmogorov–Smirnov test ($p<0.01$).
In terms of posts, Performers in total have uploaded $73,233$ photos, $43,860$ videos and $4,867$ clips, with the first two being part of their media feeds, %(requiring a subscription to access)
while the clips are sold separately. Figure~\ref{fig:posts} shows the CDF of the total number of posts. We observe that Performers earning income have clearly more posts than the ones who do not, however, the majority of Performers have less than ten postings (61\% and 93\% for the revenue and non-revenue generating ones, respectively). Again, a two-tailed Kolmogorov–Smirnov test confirms that the difference between the distributions of the number of posts for revenue and non-revenue earning Performers is significant ($p<0.01$). The low number of posts indicates that Performers, generally prefer to share their content in outlets different than FanCentro. 

\subsection{FanCentro content}

To get a better understanding of the content Performers upload in FanCentro, we analyze their media feeds which, in terms of access, can contain two kinds of posts: \textit{private} (only accessible by paying subscribers to their media feed) and \textit{public} (freely accessible). In our dataset, the majority (89\%) of posts are private ($104,737$ posts), while the rest are public ($12,356$ posts). 

\begin{figure}[!ht]
      \centering
   \includegraphics[width=1.02\columnwidth]{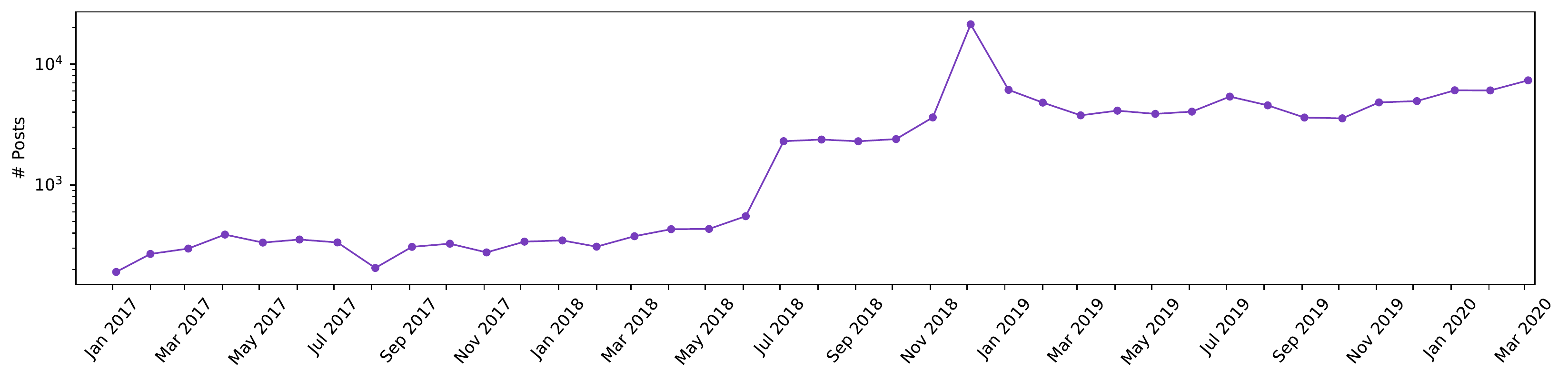}
    \caption{Monthly posting activity}
    \label{fig:monthly_posts}
\end{figure}

\begin{figure*}[!hb]
  \centering
  \begin{subfigure}[t]{0.345\textwidth}
      \centering
    \includegraphics[width=\textwidth]{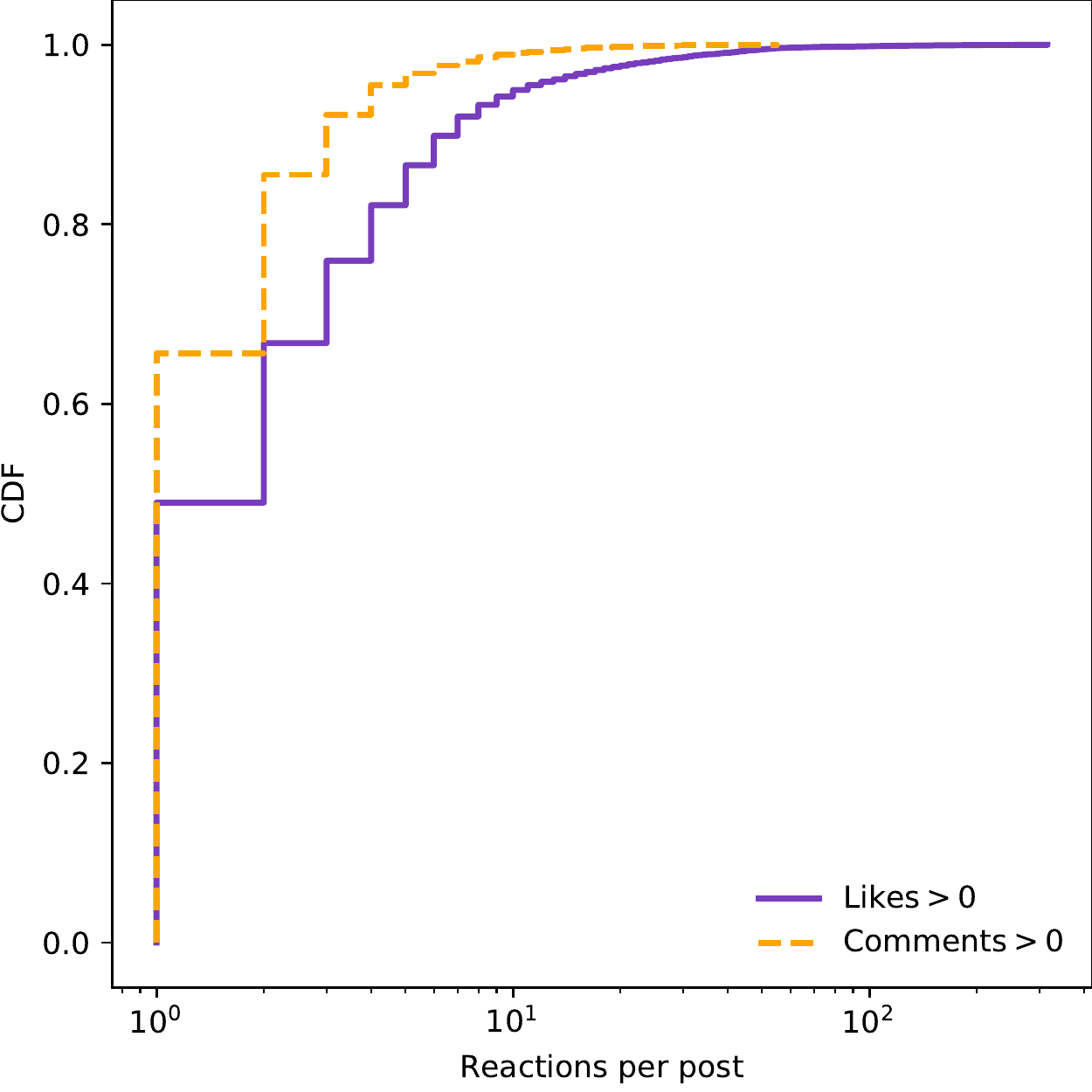}
    \caption{CDFs of likes/comments}
    \label{fig:reactions_cdf}
  \end{subfigure}
  \begin{subfigure}[t]{0.610\textwidth}
      \centering
    \includegraphics[width=\textwidth]{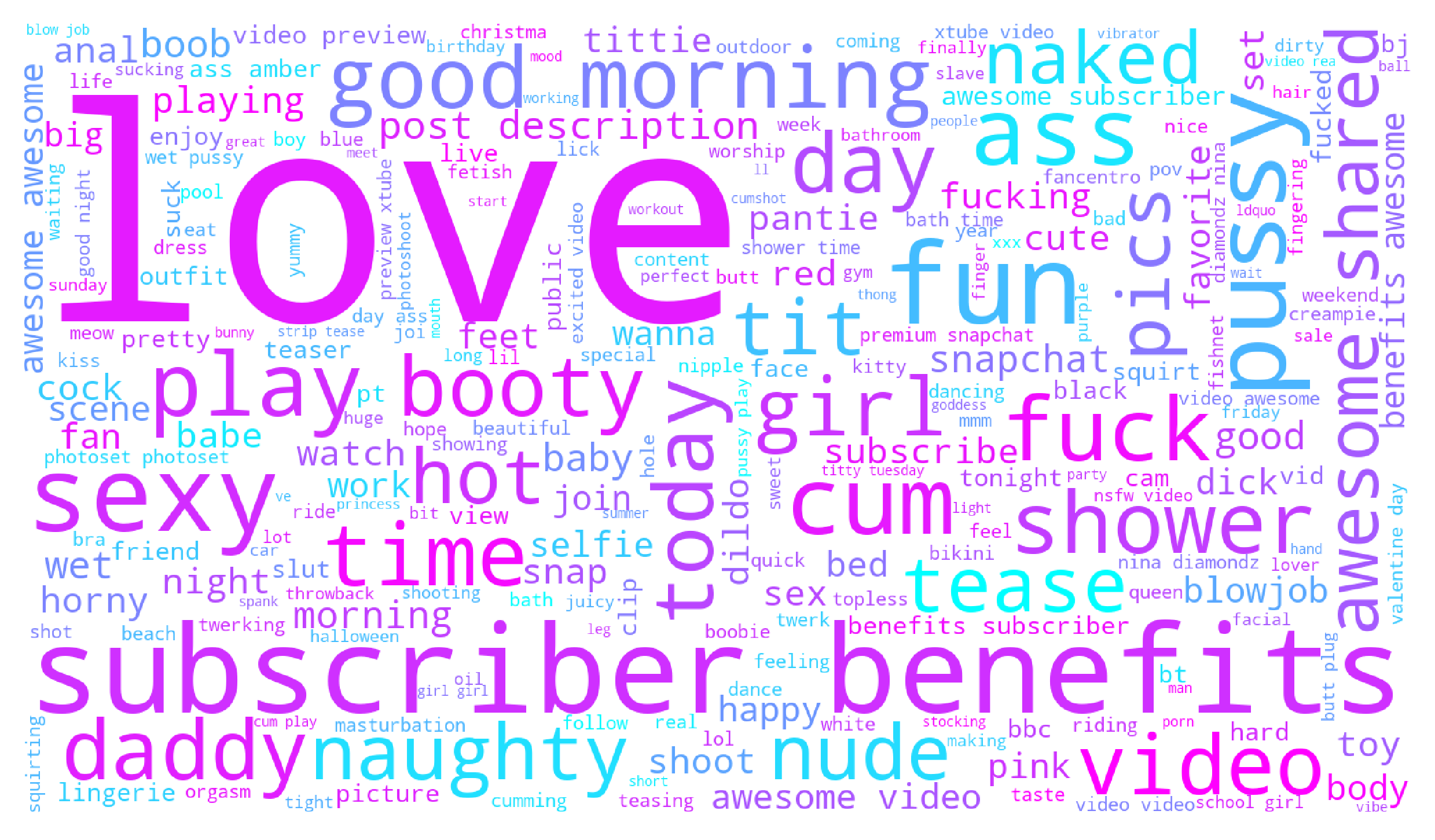}
    \caption{Word Cloud of post titles}
    \label{fig:wc_posts}
  \end{subfigure}  
  \caption{Characteristics of posts in terms of text content and reactions (likes, comments).}
  \label{fig:post_plots}
\end{figure*}

In Figure~\ref{fig:monthly_posts}, we depict the number of posts per month. We observe a consistently increasing trend in the number of posts, with a spike of $21,300$ posts in December 2018, followed by March 2020 ($7,325$ posts), which is the second most active month in terms of posting activity.
Next, we examine the characteristics of Performers' posts in terms of text content (titles) and user reactions, which results are depicted in Figure~\ref{fig:post_plots}. In our dataset, user reactions to Performers' posted content are relatively scarce, with 79\% and 92\% of the posts receiving zero likes and comments, respectively. This behaviour can be observed in Figure ~\ref{fig:reactions_cdf}, which shows the CDFs of the reactions per post. %that received likes and comments. 

Notably, the majority of these posts received just one reaction, while the most popular post in our dataset has 316 likes and 55 comments. The low number of reactions in posts comes in contrast with the relatively large numbers of followers that Performers attract, as showcased previously. In fact, there exists only a moderate correlation between the number of reactions per post and the total number of a Performer's followers (Spearman's $\rho=0.48$). In Figure~\ref{fig:wc_posts}, we created a WordCloud of the post titles. It is apparent that, apart from terms of endearment and sexual terms, the phrase ``subscriber benefits'' is prevalent, which could provide an explanation for our previous observation: to a significant extend, Performers might use FanCentro media feed posts as an additional means to promote their premium content in other channels.

Finally, we study the characteristics of the video clips uploaded by Performers, which are sold separately. The $4,867$ clips in our dataset were produced by 920 Performers, 285 (31\%) of which had non-zero revenue. A subset of $1,078$ clips is categorized as ``free for followers'', meaning that the Performers' followers can view these clips for free. This could explain the high numbers of followers that some of the Performers attract since this is a characterizing behaviour of the consumers of adult content \cite{lykousas2018adult}. Clips in our dataset have a mean duration of approximately 8 minutes and an average price of 11 USD per clip, while clip duration and price are moderately correlated (Spearman's $\rho=0.45$).

\subsection{Premium social media accounts}
We conclude our analysis by examining the different payment models for accessing the different channels used by Performers to distribute their private content. In FanCentro, the purchasable services include access to ``premium'' Snapchat and Instagram accounts and the platform's private media feed\footnote{
Recently FanCentro has introduced a purchasable direct messaging service enabling direct communication between users and Performers. Nonetheless, we excluded it from our analysis due to the low number of observations in our dataset.}.
%and a recently introduced direct messaging service enabling intimate communication between users and Performers\footnote{\url{https://www.xbiz.com/news/247941/fancentro-adds-direct-messaging-to-platform}}.
In the collected data we identified three separate payment models for accessing these services: \emph{one-time}, \emph{recurring} and \emph{free trial}.
The first two refer to one-off and recurring payments to access new content, respectively, while the ``free trial'' model allows customers to have a month of free access to the specific service, before reverting to recurring subscription payment.
In Table \ref{tab:services}, we present the distribution of the different payment models for the offered services. Private Snapchat is by far the most popular premium service, and the majority of Performers prefer offering their services as subscriptions.
\begin{table*}[!ht]
\centering
\begin{tabular}{c c c c  >{\columncolor[gray]{0.95}}r }
    \toprule
    \multirow{2}[3]{*}{\thead{Premium\\Service}} & \multicolumn{4}{c}{\thead{Payment model}}\\ 
    \cmidrule(rl){2-5}
    & Reccuring  & One-time & Free Trial & Total  \\ 
    \midrule
    \multicolumn{1}{l}{Snapchat}& 11635 & 1153 & 41 & 12829   \\
    \multicolumn{1}{l}{FanCentro}& 4716 & 0 & 5 & 4721  \\ 
    \multicolumn{1}{l}{Instagram}& 1741 & 191 & 0 & 1932  \\
    \midrule
    \multicolumn{1}{l}{\textbf{Total}} & 18092 & 1344 & 46 & 19482  \\
    \bottomrule
\end{tabular}
\caption{Premium services}
\label{tab:services}
\end{table*}

\begin{figure}[!ht]
      \centering
   \includegraphics[width=1\columnwidth]{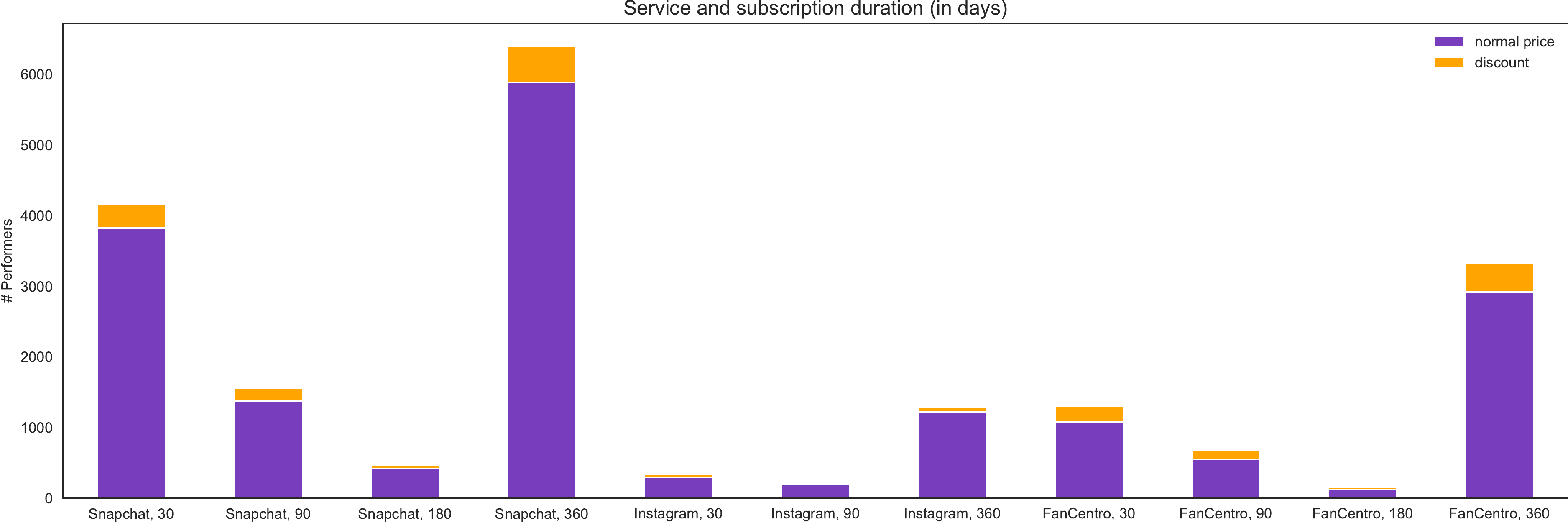}
    \caption{Number of offerings per service and subscription duration.}
    \label{fig:subscriptions_counts}
\end{figure}

\begin{figure*}[!hb]
  \centering
  \begin{subfigure}[t]{0.325\textwidth}
      \centering
    \includegraphics[width=\textwidth]{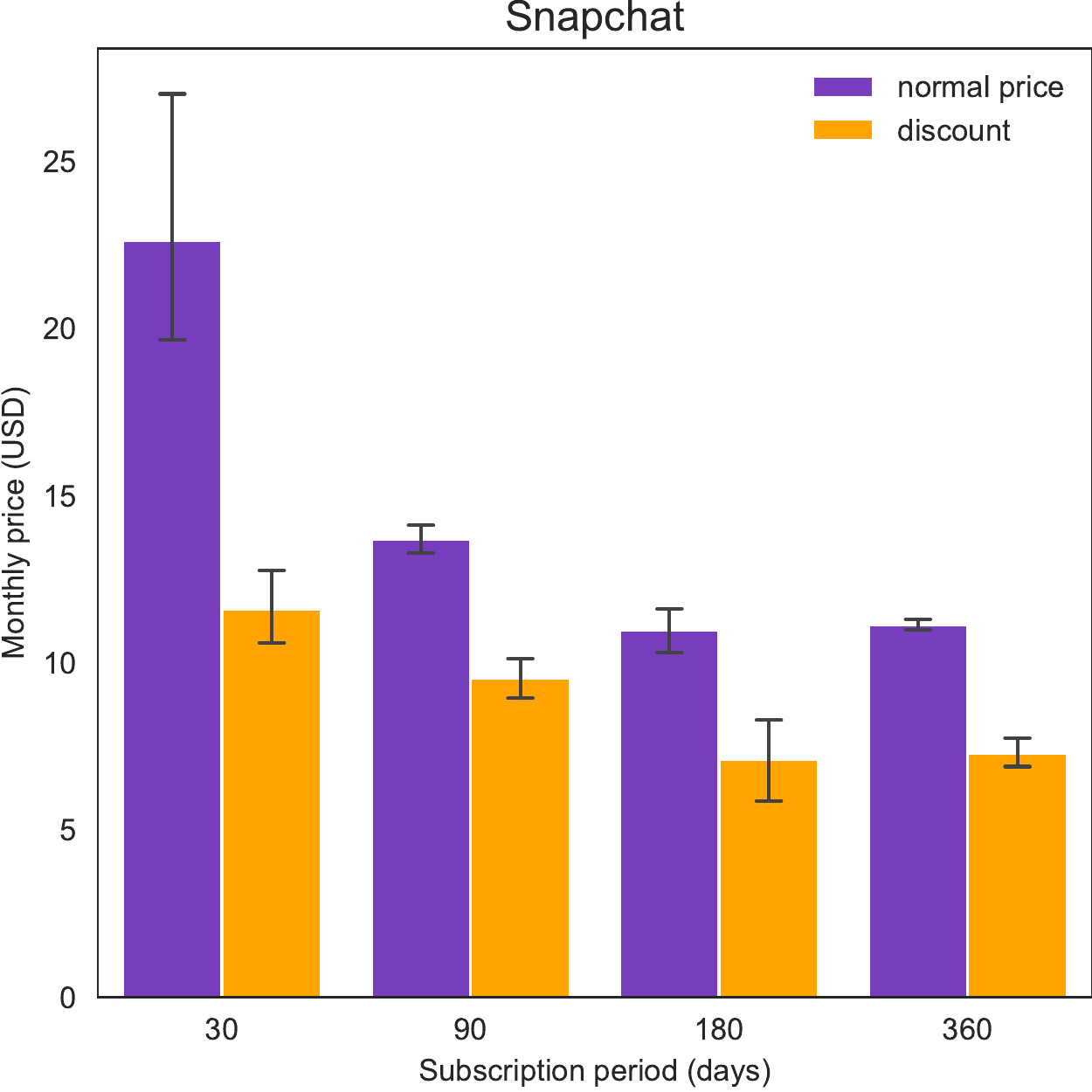}
    \caption{Snapchat}
    \label{fig:snap_price}
  \end{subfigure}
  \begin{subfigure}[t]{0.325\textwidth}
      \centering
    \includegraphics[width=\textwidth]{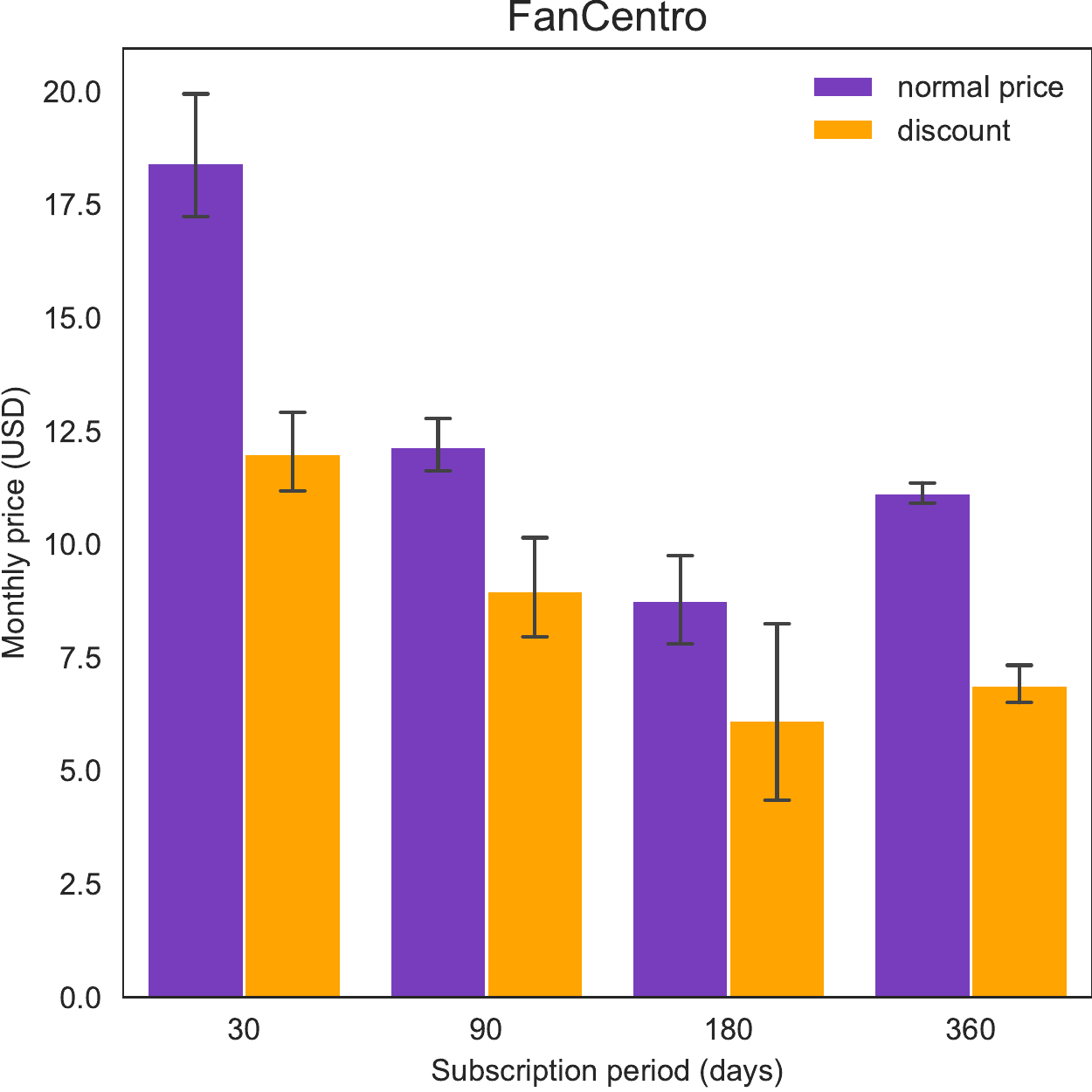}
    \caption{FanCentro}
    \label{fig:fancentro_price}
  \end{subfigure}  
  \begin{subfigure}[t]{0.325\textwidth}
      \centering
    \includegraphics[width=\textwidth]{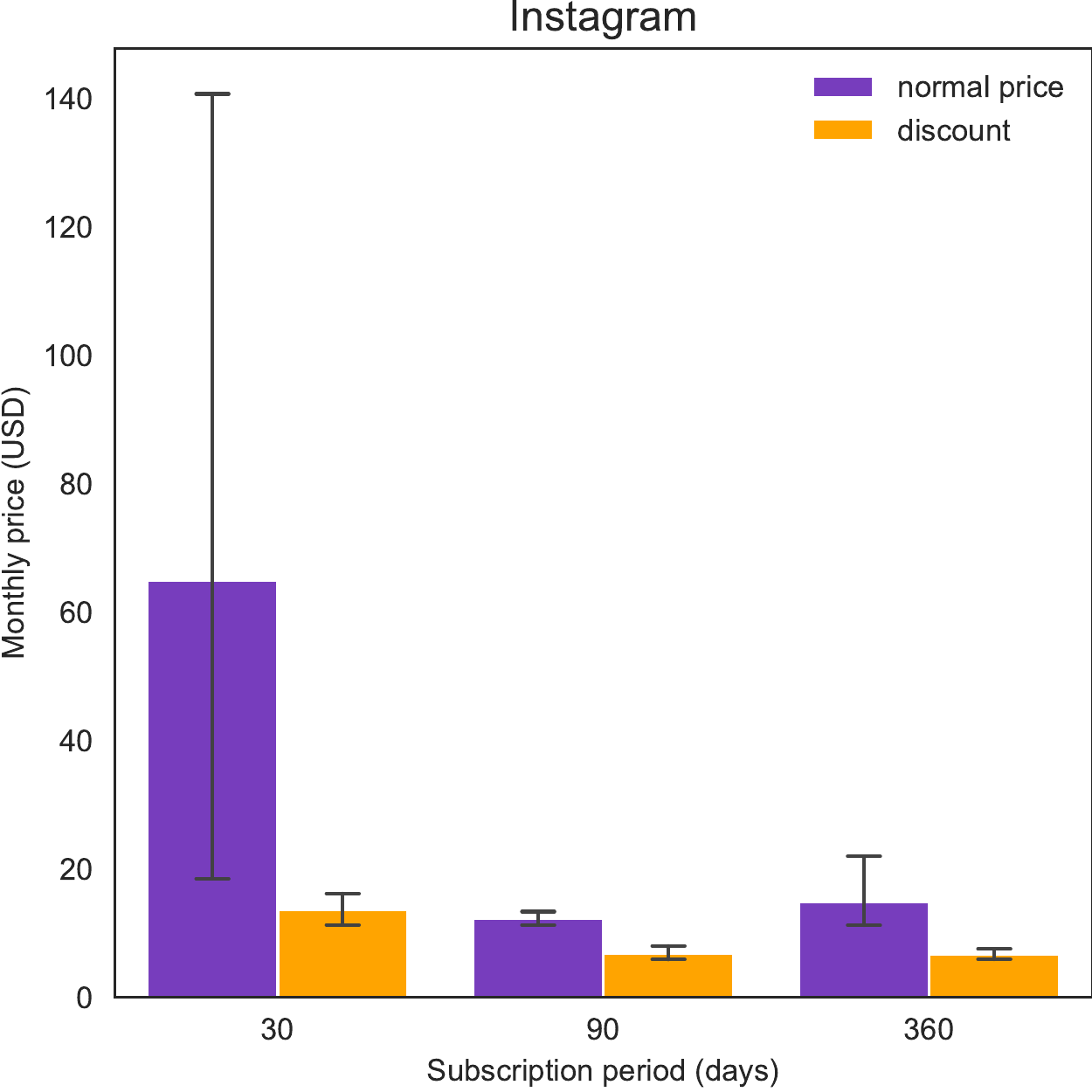}
    \caption{Instagram}
    \label{fig:instagram_price}
  \end{subfigure}  
  \caption{Bar plots of monthly subscription price (normal and discounted) per service and subscription duration.}
  \label{fig:barplots}
\end{figure*}

%reccuring
% the premium services offered, we focus on the recurring sub
The mean price of the Performers selling their services under one-off payments is 30 USD %(std=66)
for Snapchat and 32 USD %(std=50) 
for Instagram.
To get a deeper insight into the recurring payment model adopted by the majority of Performers, in Figure~\ref{fig:subscriptions_counts} 
we present the distribution of subscription offerings, and in Figure~\ref{fig:barplots} we show the monthly subscription price distribution per service and total subscription duration. For simplicity, we only consider the subscription periods with more than 100 occurrences in our dataset. We note that Performers can offer their services at discounted rates as a means of promotion (similar to free trial access), which comprise a small fraction of the total offerings ($2,004$ in total).

In Figure~\ref{fig:subscriptions_counts}, we observe that the most popular service is the yearly Snapchat subscription, offered by $5,892$ performers, followed by monthly Snapchat subscription ($3,828$ offerings) and yearly access to FanCentro feed ($2,921$ offerings). While three-month and half-year subscriptions exist, they are not common, accounting only for 25\% of total offerings. %they are actually 4256/16488
The subscription fee is calculated on the total subscription period. As such, the monthly price generally decreases as the subscription duration increases. The monthly subscription to Performers' premium accounts, which is the pricier option in all cases, on average costs 21.7 USD %(std=112) 
and 58 USD %(std=513)
for Snapchat (Figure~\ref{fig:snap_price}) and Instagram (Figure~\ref{fig:instagram_price}) accounts, respectively. 
Notably, in the first case, the price can go up to $5,000$ USD, and in the second case up to $8,000$ USD. In this regard, the lowest priced service is access to FanCentro media feed ($\mu=17.3$ USD), % \sigma=20$), 
which can cost up to 500 USD monthly (Figure~\ref{fig:fancentro_price}).
Nevertheless, the most common subscription duration is one year, priced on average 10 USD/month for Snapchat and FanCentro feed, and 14 USD/month for Instagram. Additionally, discounted rates show an average decrease of 6 USD/month for Snapchat, 14 USD/month for Instagram and 4 USD/month for FanCentro feed, when compared to the normal prices of each service, respectively.

% \begin{figure*}[!ht]
% \begin{minipage}{.325\linewidth}
% \centering
% \subfloat[]{\label{fig:snap_price}\includegraphics[width=\textwidth]{Snapchat_price.pdf}}
% \end{minipage}%
% \begin{minipage}{.325\linewidth}
% \centering
% \subfloat[]{\label{fig:fancentro_price}\includegraphics[width=\textwidth]{FanCentro_price.pdf}}
% \end{minipage}%
% \begin{minipage}{.325\linewidth}
% \centering
% \subfloat[]{\label{fig:instagram_price}\includegraphics[width=\textwidth]{Instagram_price.pdf}}
% \end{minipage}\par\medskip
% \centering
% \subfloat[]{\label{fig:subscriptions_counts}\includegraphics[width=\textwidth]{subscriptions.pdf}}
% \label{fig:subscriptions}
% \caption{Bar plots of monthly subscription price (normal and discounted) per service and subscription duration, and number of offerings per service and subscription duration,}
% \end{figure*}

% \todo[inline]{FC what you do below is directly a conclusions. Whether you extract part of the experiments section and discuss it here,  or you name this last section below discussion and conclusions. In general, I believe that discussion section should have enough entity by itself.}

\section{Conclusions}
In this work, we performed the first quantitative analysis of the semi-illicit adult content market layered on the top of popular social media platforms like Snapchat and Instagram. To this end, we studied the demographics and activity of the selling users in FanCentro, as well as some descriptive characteristics of the content they upload. The existence of sites like FanCentro where Performers can openly sell and promote premium social media accounts indicate that the industry built on the inefficacy of social media platforms to enforce community guidelines for effectively banning adult content is here to stay. This inefficacy is exploited and monetized in large scale, exacerbated by the fact the explicit content is staying ``hidden'' in private accounts, access to which is sold through the different models studied. 

Moreover, our findings indicate that the coronavirus-induced lockdowns have accelerated the growth of this marketplace. 
This phenomenon is also reflected by the rise of other influencer-centric adult content markets, such as OnlyFans, which observed a major increase in traffic during the coronavirus pandemic \cite{nytimes2020sexwork,wired2020corona}. In part, this is due to the fact that a large number of sex workers lost their original revenue streams because of the virus; in addition, an increasing number of influencers transition to online sex work as a means to adapt to the economic downturn which caused companies to reduce marketing budgets, that would have been otherwise used for sponsored content \cite{elle2020onlyfans}.
The strong online presence of Performers across multiple popular social media sites where they openly promote their paid content signals the shift of online adult content industry towards an increasingly mainstream, gig economy. %influencer-based,
Nonetheless, the proliferation of adult content flowing unobstructed through social media, diffused and being promoted via users with large followings, might pose a serious threat to the safety of mainstream online communities, especially for the younger users.

\section*{Acknowledgements}
This work was supported by the European Commission under the Horizon 2020 Programme (H2020), as part of the project \textit{LOCARD} (\url{https://locard.eu}) (Grant Agreement no. 832735).

The content of this article does not reflect the official opinion of the European Union. Responsibility for the information and views expressed therein lies entirely with the authors.

The authors would also like to thank the anonymous reviewers for their valuable comments and suggestions.
%indicate that industry is built on the inefficacy of popular social media platforms to enforce community guideline prohibiting adult content effectively. This inefficacy is exploited and monetized on a large scale since the explicit content is staying "hidden" in private accounts, access to which is sold and marketed by Performers in plain sight.

\bibliographystyle{splncs04}
\bibliography{references}
\end{document}